\pdfoutput=1

\documentclass[preprintnumbers,showpacs,floats,twocolumn,prd,aps]{revtex4}
\usepackage{amssymb,amsmath}
\usepackage{epsfig,hyperref}

\makeatletter
\renewcommand{\vec}[1]{\mbox{\boldmath$#1$}}
\newcommand{\arXiv}[2][]{\href{http://arxiv.org/abs/#2}{\texttt{arXiv:#2\@ifempty{#1}{}{ [#1]}}}}
\makeatother

\begin{document}

\title{Critical Collapse and Solitons in Classical Conformal Field Theory}
\author{Andrei V. Frolov}\email{frolov@sfu.ca}
\affiliation{
  Department of Physics,
  Simon Fraser University\\
  8888 University Drive,
  Burnaby, BC Canada
  V5A 1S6
}
\date{October 13, 2011}

\begin{abstract}
  We study the fate of a localized wavepacket in a classical conformal field theory with attractive interaction $V(\phi) = -\frac{1}{4}\, \lambda\phi^4$. As potential is unbounded from below, homogeneous field collapses to singularity in finite time. However, finite size wavepacket can disperse before it collapses. Competition between the two outcomes results in a critical behavior, much like the one seen in gravitational collapse. We calculate the critical exponents, and show that there are static regular soliton-like solutions in the theory.
\end{abstract}

\preprint{SCG-2011-06}
\maketitle

\section{Introduction}

Scale-invariance is ubiquitous in physics, appearing in statistical mechanics and phase transitions, quantum field theories near fixed point, and the much-lauded AdS/CFT correspondence \cite{Aharony:1999ti}. In this paper, we explore dynamics of a simple classical conformal field theory in a flat spacetime, consisting of a single real-valued scalar field with action
\begin{equation}\label{eq:action}
  S = \int \left[-\frac{1}{2} (\nabla\phi)^2 - V(\phi) \right] \sqrt{-g}\, d^4 x
\end{equation}
and the quartic potential with the ``wrong'' sign
\begin{equation}\label{eq:V}
  V(\phi) = -\frac{1}{4}\, \lambda\phi^4,
\end{equation}
corresponding to attractive self-interaction. Although potential is unbounded from below and the theory is unstable, the classical vacuum $\phi=0$ is infinitely long-lived, and the corresponding quantum theory is renormalizable \cite{Symanzik:1973hx, Parisi:1973ma}. In fact, this model is in some sense better off than familiar gravity, which is also attractive, also has unbounded negative binding energy, singularities appearing in evolution of a regular initial data, but is non-renormalizable on top.

Quantum field theory with $\phi^4$ interaction has been extensively studied, with renormalization group calculated perturbatively to fifth loop long time ago \cite{Kleinert:1991rg}. Recently, Dvali {\em et.~al.} \cite{Dvali:2011uu} argued that the theory (\ref{eq:action},\ref{eq:V}) captures some features of quantum phenomena, asymptotic freedom and dimensional transmutation, at a classical level. This would be pretty interesting, because in classical field theory we have direct access to fully non-perturbative and non-linear solutions, which are readily found using numerical methods. Models of the type (\ref{eq:V}) might also have cosmological applications, where conformal symmetry can be used to generate scale-invariant primordial fluctuations \cite{Rubakov:2009np, Libanov:2010nk}.

With this in mind, it would seem that the theory (\ref{eq:V}) warrants some further investigation. Seeming simplicity of the model is deceptive, and there are some rather non-trivial things about it, which is what I report here.

Evolution of the classical scalar field (\ref{eq:action}) with potential (\ref{eq:V}) is governed by the wave equation
\begin{equation}\label{eq:wave}
  \Box\phi + \lambda\phi^3 = 0.
\end{equation}
Conformal symmetry can be used to make variables dimensionless by rescaling to an arbitrary length scale $\ell$. In addition coupling $\lambda$ can be rescaled to unity by 
\begin{equation}\label{eq:scaling}
  \bar{\phi} = \ell \phi, \hspace{1em}
  \bar{x}^\mu = \lambda^{\frac{1}{2}} \ell^{-1}\, x^\mu,
\end{equation}
as long as it does not vanish, which is what we will do in most calculations below.

The outline of the paper is as follows: In Section~\ref{sec:collapse}, I explore the eventual fate of a localized wavepacket and find that there is dynamical critical behavior in this theory, similar to the critical gravitational collapse \cite{Choptuik:1992jv}. In Section~\ref{sec:static}, I derive static spherically symmetric solutions, which turn out to have regular localized soliton-like states that are unstable and probably correspond to critical solutions of the dynamical collapse. As the energy of these is infinite due to slowly decaying tails, I construct static non-linear dipole configurations in Section~\ref{sec:dipole}, and show that their energy is finite. I conclude with some speculations in Section~\ref{sec:discussion}. Finer points of classical beta function asymptotics and behavior of irregular static solutions are hidden in Appendix~\ref{sec:running}.

\section{Critical Collapse}
\label{sec:collapse}

\begin{figure*}[t!]
  \hspace{-18pt}
  \begin{tabular}{ccc}
    \epsfig{file=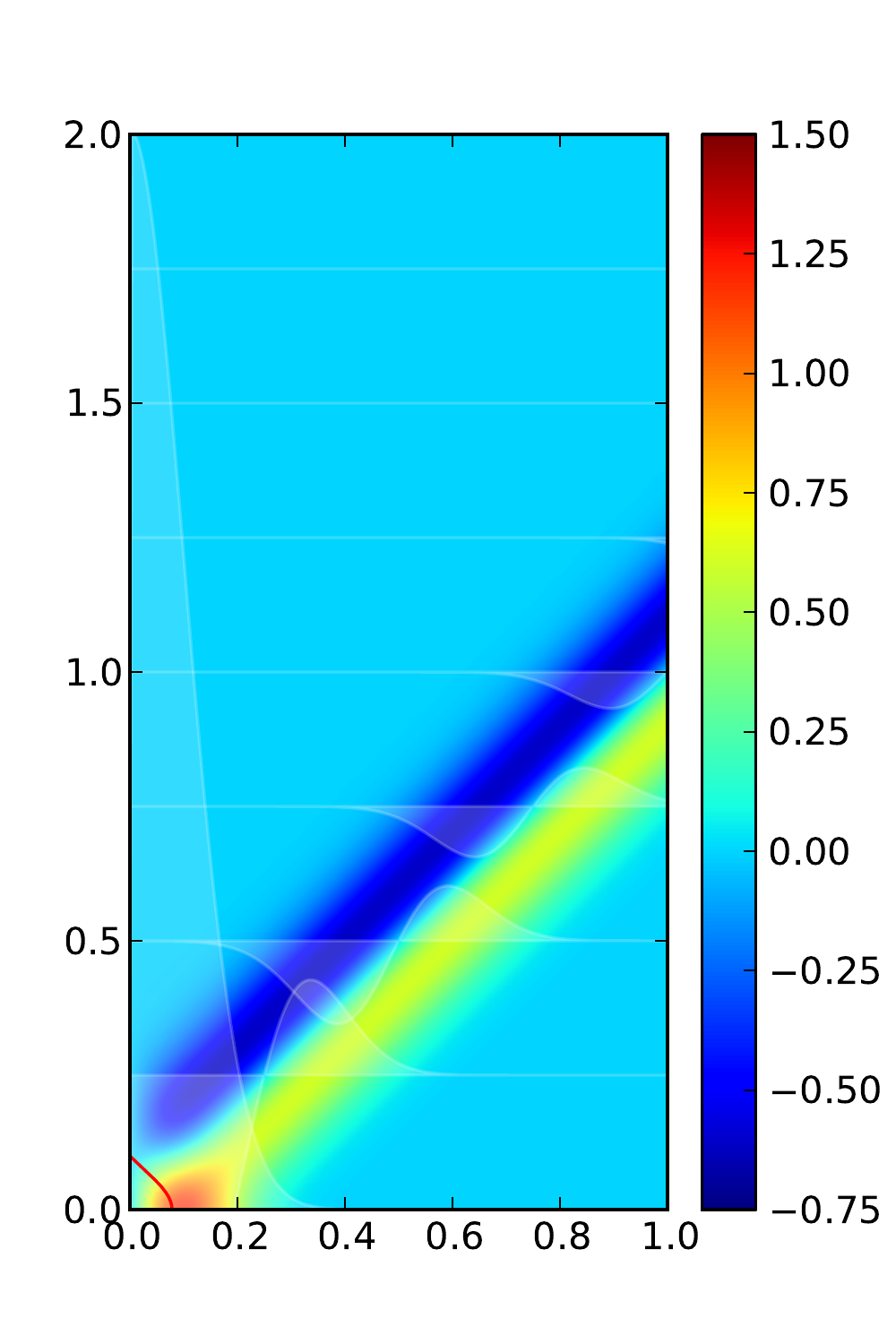, width=6cm} &
    \epsfig{file=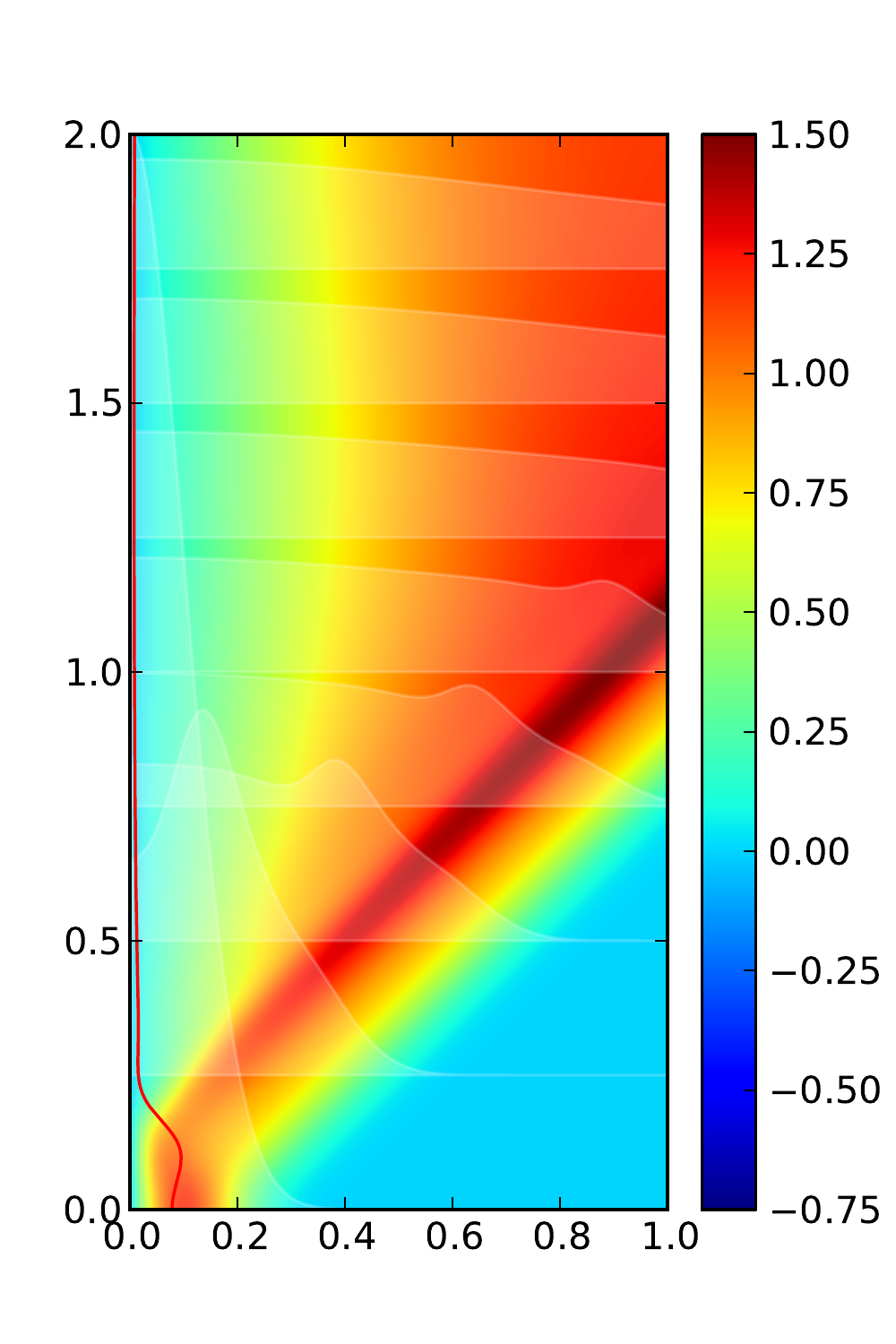, width=6cm} &
    \epsfig{file=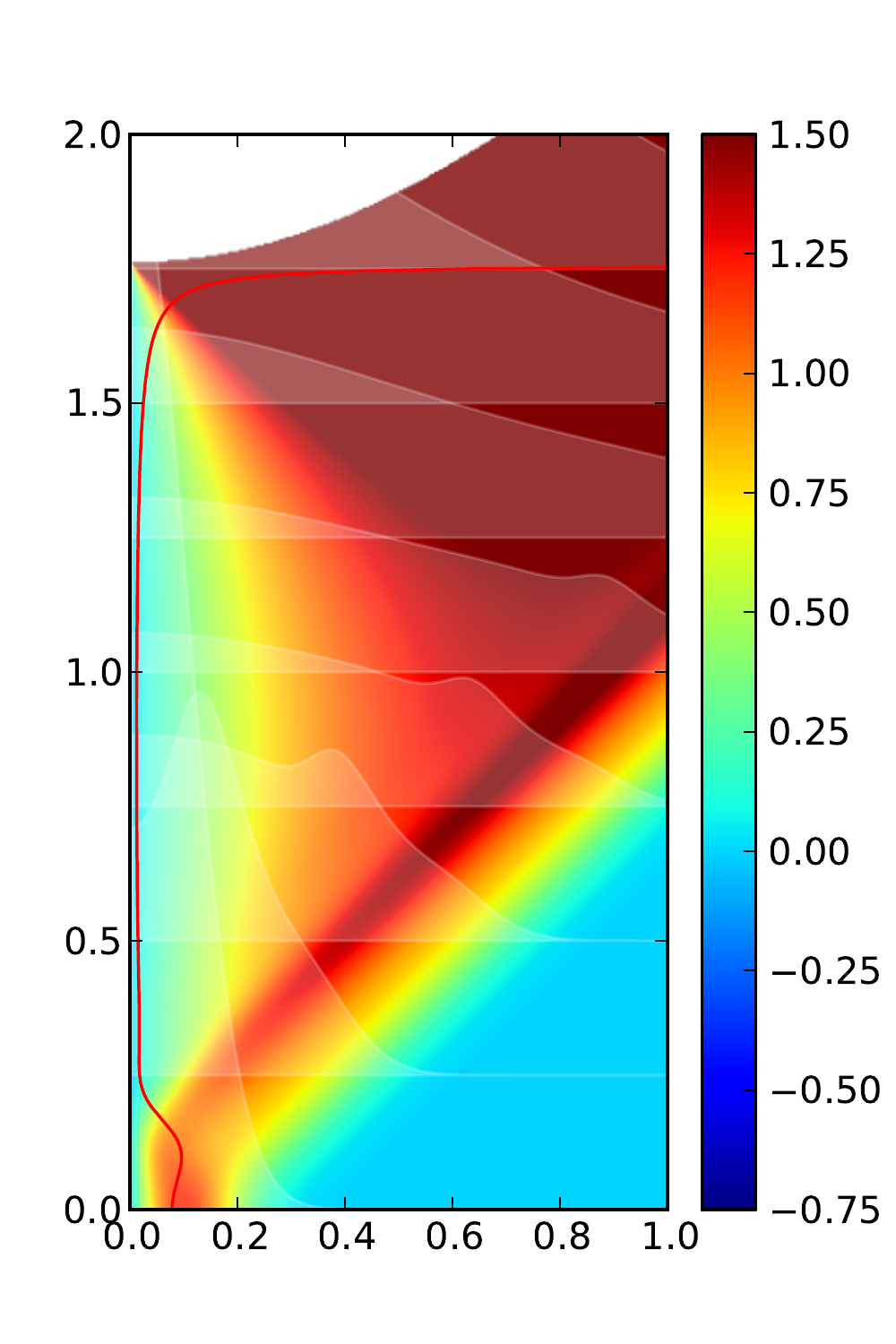, width=6cm} \\
    (a) & (b) & (c) \\
  \end{tabular}
  \caption{Dispersal of a momentarily stationary spherical Gaussian wavepacket in free theory (a), and in non-linear CFT with initial amplitude slightly below (b) and slightly above (c) critical value. In the latter case, singularity forms in finite time. Density plot in $\{r,t\}$-plane shows isosurfaces of $q(r,t) \equiv r\phi$, with evolution of field profiles $\phi(r,t=\text{const})$ overlaid in pale shading. Solid red line plots field value at origin $\phi(r=0,t)$ as a function of time.}
  \label{fig:disperse}
\end{figure*}
\begin{figure*}[t!]
  \hspace{-18pt}
  \begin{tabular}{ccc}
    \epsfig{file=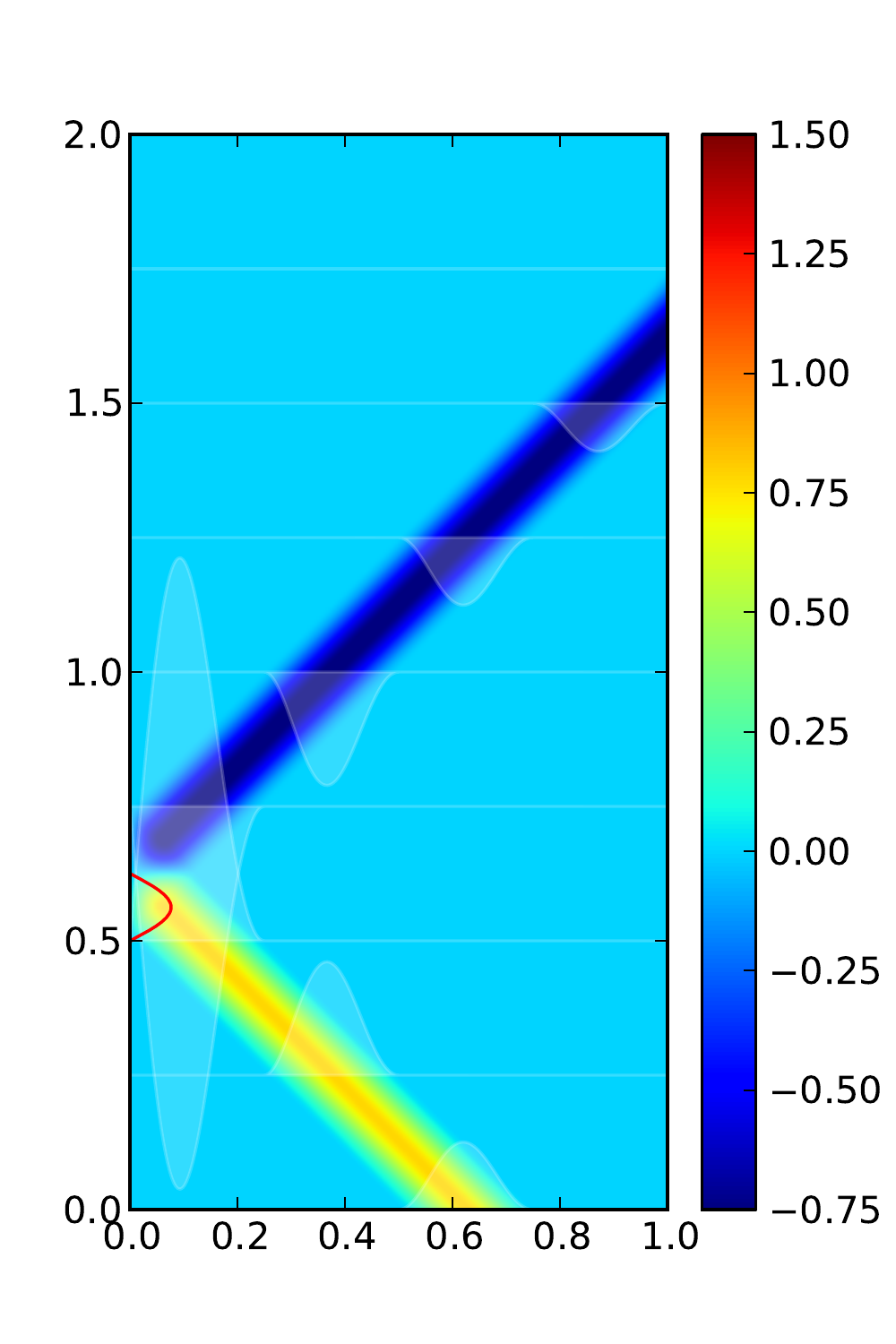, width=6cm} &
    \epsfig{file=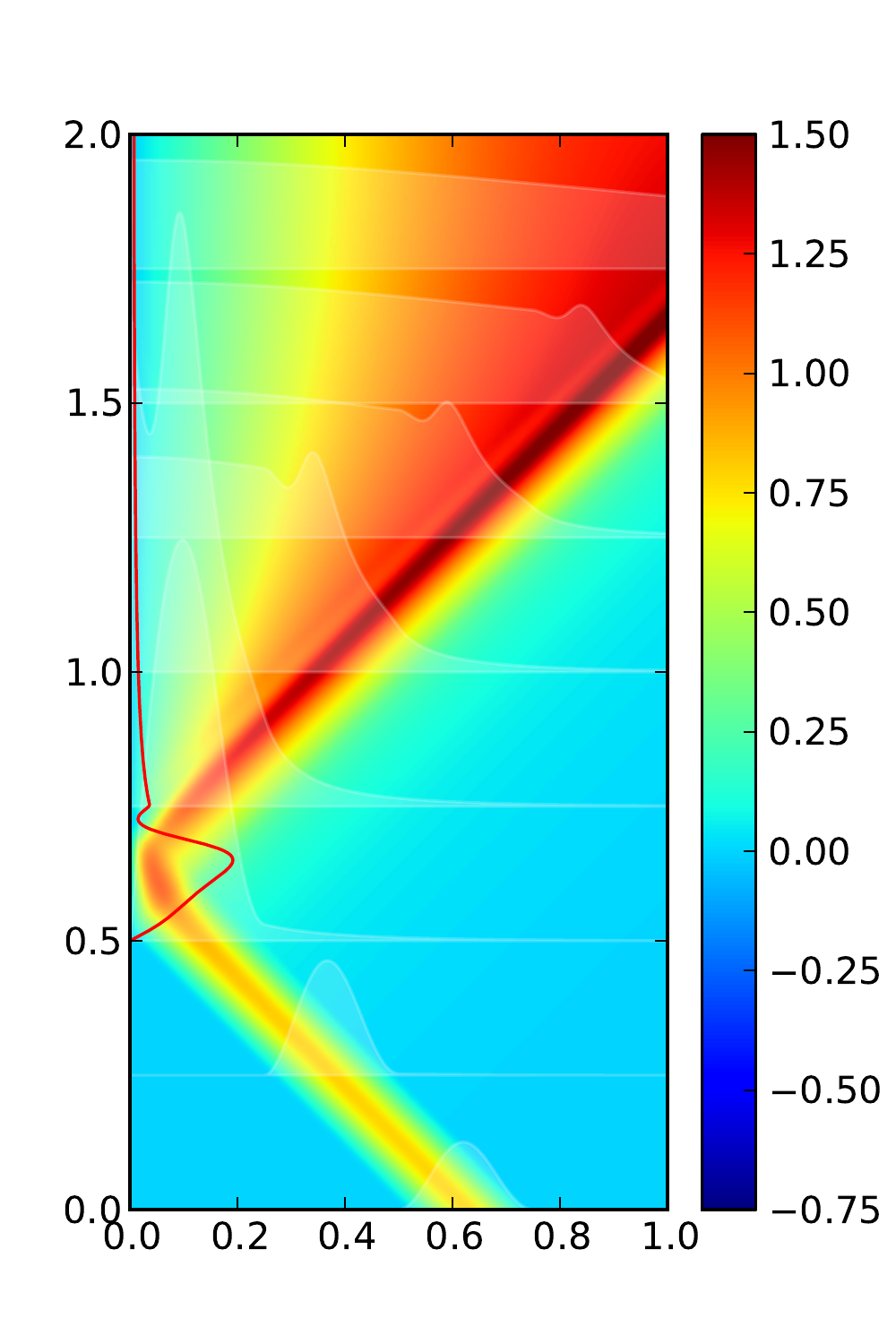, width=6cm} &
    \epsfig{file=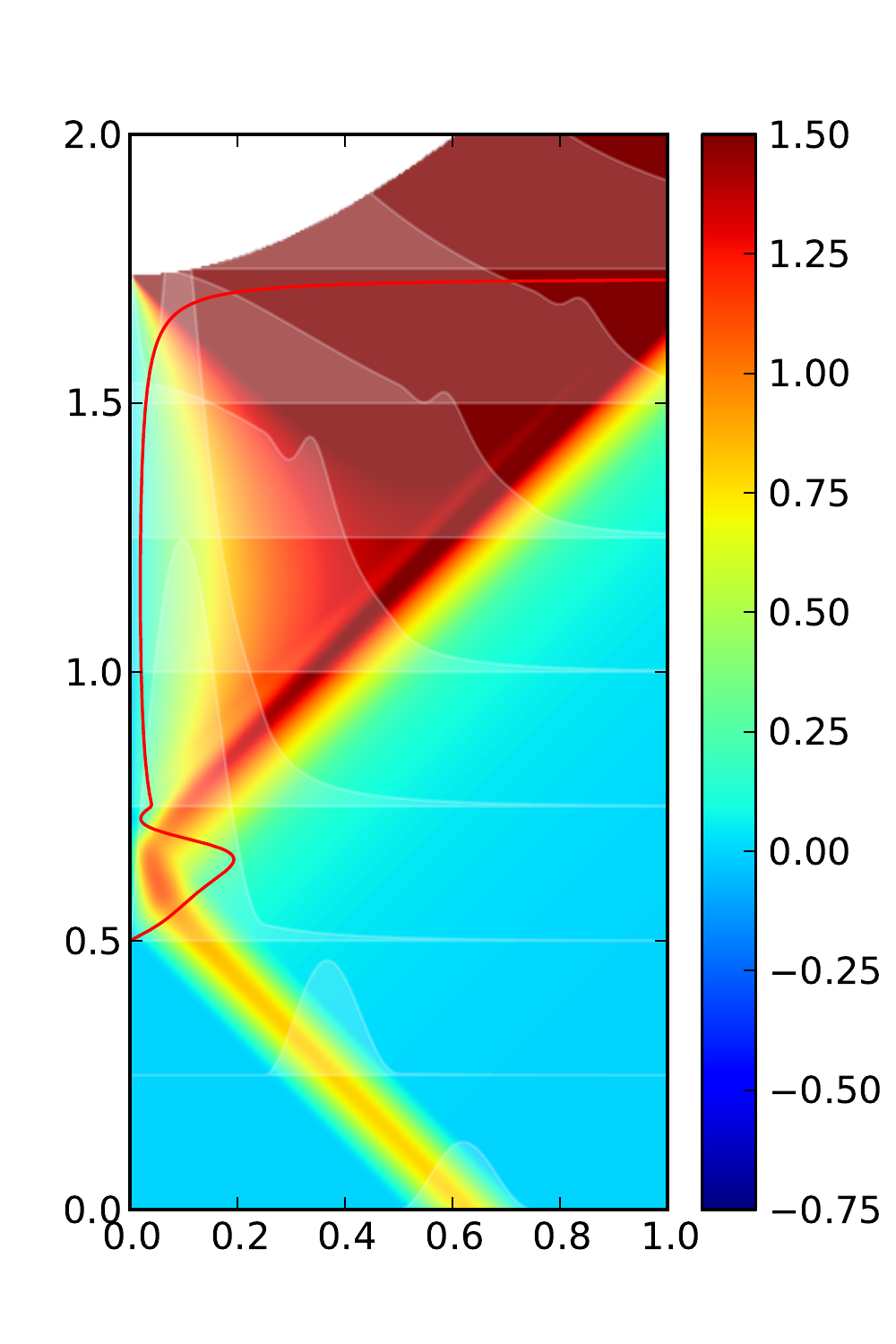, width=6cm} \\
    (a) & (b) & (c) \\
  \end{tabular}
  \caption{Scattering of an ingoing sine squared wavepacket in free theory (a), and in non-linear CFT with initial amplitude slightly below (b) and slightly above (c) critical value. In the latter case, singularity forms in finite time. \protect{Same legend as Fig.~\ref{fig:disperse}.}}
  \label{fig:scatter}
\end{figure*}

Conformally invariant theory with negative potential will have a direction in which the potential is unbounded from below, and hence is unstable. This is not necessarily fatal by itself. As long as the vacuum does not decay explosively, this instability simply means that fields are prone to collapse under self-attraction, just like Jeans instability in gravitating matter. Just as in gravity, singularities can form as a result of a field collapse in an attractive CFT. 

So, what is the eventual fate of a given field distribution in an attractive CFT?

\begin{figure*}
  \begin{tabular}{cc}
    \epsfig{file=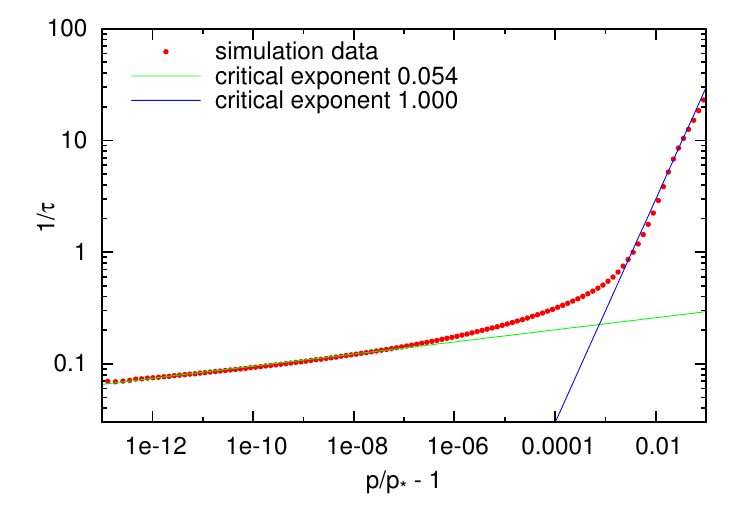, width=8cm} &
    \epsfig{file=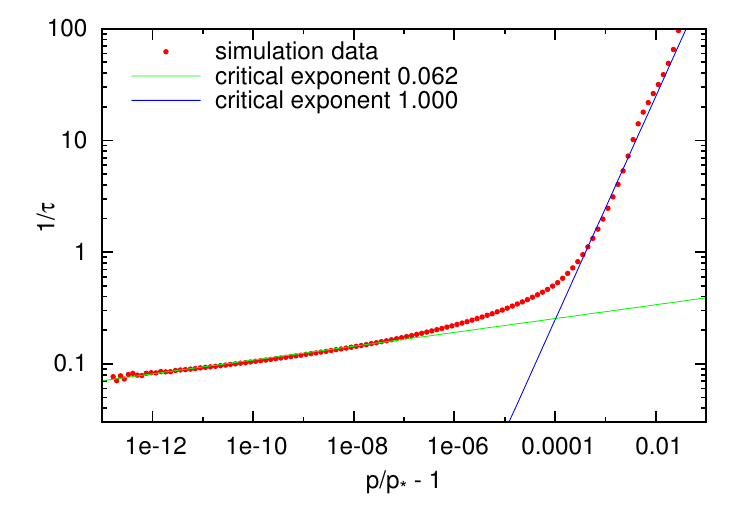, width=8cm}\\
    (a) & (b) \\
  \end{tabular}
  \caption{Scaling of lifetime $\tau$ of momentarily stationary (a) and scattered (b) wavepacket with initial field amplitude $p$.}
  \label{fig:scaling}
\end{figure*}

Dynamics of the homogeneous field evolution are the same as for a particle falling down in potential (\ref{eq:V}); a particular solution of (\ref{eq:wave}) is
\begin{equation}\label{eq:t*}
  \phi(t) = \frac{\phi_0}{1 - t/t_*}, \hspace{1em}
  t_* = \sqrt{\frac{2}{\lambda}}\,\frac{1}{\phi_0}.
\end{equation}
The field collapses from finite value $\phi_0$ to singularity in a finite time $t_*$, with the timescale set by the field value. The vacuum solution, although unstable, takes infinitely long to decay as the potential is rather flat at the top.

Localized wavepacket is not going to hang around waiting for the collapse to happen. Characteristics of the wave equation (\ref{eq:wave}) propagate at the speed of light, so the initially stationary wavepacket will disperse in a time roughly comparable to its size. Competition between propagation and self-attraction results in a critical behavior much like the one seen in gravitational collapse \cite{Choptuik:1992jv, Koike:1995jm, Gundlach:2007gc}, but in a much simpler theory.

To analyze what happens, lets solve the equation (\ref{eq:wave}) for a spherical wave
\begin{equation}
  - \frac{\partial^2\phi}{\partial t^2} + \frac{\partial^2\phi}{\partial r^2} + \frac{2}{r}\,\frac{\partial\phi}{\partial r} + \lambda\phi^3 = 0.
\end{equation}
With the usual variable change
$\phi(r,t) = q(r,t)/r$,
it reduces to a one dimensional wave equation
\begin{equation}
  - \frac{\partial^2 q}{\partial t^2} + \frac{\partial^2 q}{\partial r^2} + \lambda\, \frac{q^3}{r^2} = 0
\end{equation}
with Dirichlet boundary condition at the origin $r=0$
\begin{equation}
  q = 0\Big|_{r=0}, \hspace{1em}
  \frac{\partial q}{\partial r} + \frac{\partial q}{\partial t} = 0\Big|_{r=r_c},
\end{equation}
supplemented by outgoing boundary condition at outer domain boundary $r=r_c$ for simulation. The semi-linear partial differential equation is straightforward to solve numerically; we use second order leapfrog method on a large uniform grid ($2^{15}$ nodes for results quoted here).

Evolution of initially stationary Gaussian wavepacket $\phi(r, t=0) = p \exp(-r^2/s^2)$ is shown in Fig.~\ref{fig:disperse}. In free theory (\ref{fig:disperse}a), momentarily stationary configuration splits into a left and right moving waves, both of which disperse to infinity after the left mover reflects off of origin. For weak non-linearity, the evolution is pretty much the same, but as one increases the amplitude $p$ (keeping $s$ constant), situation changes. The wavefront comes out without the phase reversal, leaving a blob of scalar field behind, which is held by attractive interaction. If the amplitude is slightly smaller than a certain value $p_*$, it hangs close to balance but disperses eventually as in Fig.~\ref{fig:disperse}b. If one increases the amplitude even slightly above critical value $p_*$, singularity forms as in Fig.~\ref{fig:disperse}c. The change is very abrupt, the difference of initial amplitudes between Figs.~\ref{fig:disperse}b and \ref{fig:disperse}c is roughly $0.1\%$. Same thing happens in a scattering of initially ingoing wave, as shown in Fig.~\ref{fig:scatter}.

This is a by-now-familiar picture of the critical gravitational collapse \cite{Choptuik:1992jv, Koike:1995jm, Gundlach:2007gc}, except it happens in a pure field theory in a flat spacetime. One can take inverse lifetime $1/\tau$ (measured from the moment the maximum of the pulse arrives to the origin to the moment singularity first forms) as an order parameter for the phase transition. It is zero in a regular phase, and scales with amplitude as shown in Fig.~\ref{fig:scaling} in a singular phase. The scaling is a broken power law: For large amplitudes scaling is linear as characteristic of homogeneous collapse (\ref{eq:t*}), but close to criticality it switches over to much smaller exponent with a numerical value of approximately $1/16$, which is sufficiently close for different families of solutions to suspect some sort of universality. If the scaling persists as one tunes the parameter to criticality, the critical solution separating the two phases would be infinitely long-lived, i.e.\ quasi-static. This is unachievable in collapse simulation, but is easy to find directly if it exists, which we will do next.
\clearpage

\section{Static Solutions}
\label{sec:static}

As we have seen in the last section, arbitrarily long-lived configurations can be obtained in the field collapse by fine-tuning a single parameter. In fact, the theory admits one-parameter family of regular soliton-like solutions. Spherically symmetric static solutions satisfy
\begin{equation}\label{eq:static:phi}
  \frac{d^2\phi}{dr^2} + \frac{2}{r}\,\frac{d\phi}{dr} + \lambda\phi^3 = 0.
\end{equation}
Regularity demands that the power series expansion of the solution around the origin is an even function of $r$, while scaling symmetry guarantees that variables enter the expansion only in a combination $\xi = \lambda\phi_0^2r^2$
\begin{equation}\label{eq:static:regular}
  \phi(r) \simeq \phi_0\left[1 - \frac{\xi}{6} + \frac{\xi^2}{40} - \frac{19\,\xi^3}{5040} + \ldots\right].
\end{equation}
Unlike free theory, solutions with $\lambda > 0$ will decay at infinity regardless of the value $\phi_0$ field takes at the origin. The easiest way to see this is to do a variable change
\begin{equation}\label{eq:static:psi}
  \phi(r) = \frac{\psi(r)}{(3r)^{\frac{2}{3}}}, \hspace{1em}
  3r = x^3,
\end{equation}
which casts the equation (\ref{eq:static:phi}) into the form of time-dependent anharmonic oscillator
\begin{equation}
  \psi'' - \frac{2\,}{x^2}\, \psi + \lambda\psi^3 = 0.
\end{equation}
The above equation corresponds to canonical equations of motion for an ($x$-dependent) Hamiltonian
\begin{equation}\label{eq:static:H}
  {\cal H}(\psi,\pi;x) = \frac{1}{2}\,\pi^2 - \frac{\psi^2}{x^2} + \frac{1}{4}\, \lambda \psi^4,
\end{equation}
which makes it clear that any solution will go to a limit cycle at large $x$. Scaling $\lambda$ and amplitude of oscillations to unity, the limit cycle solution is given in terms of Jacobi elliptical cosine function, or its harmonic expansion \cite{Kiper:1984}
\begin{equation}\label{eq:cn}
  \psi(x) = \text{cn}(x,2^{-\frac{1}{2}}) = 2^{\frac{3}{2}} k \sum\limits_{n=1}^{\infty} \frac{\cos2(n-\frac{1}{2})kx}{\cosh(n-\frac{1}{2})\pi}.
\end{equation}
It is periodic with period $2\pi/k=\pi^{-\frac{1}{2}}\Gamma^2(\frac{1}{4})$, and the harmonic expansion is exponentially converging, so only a few terms are needed to accurately
represent its shape.

\begin{figure}
  \centerline{\epsfig{file=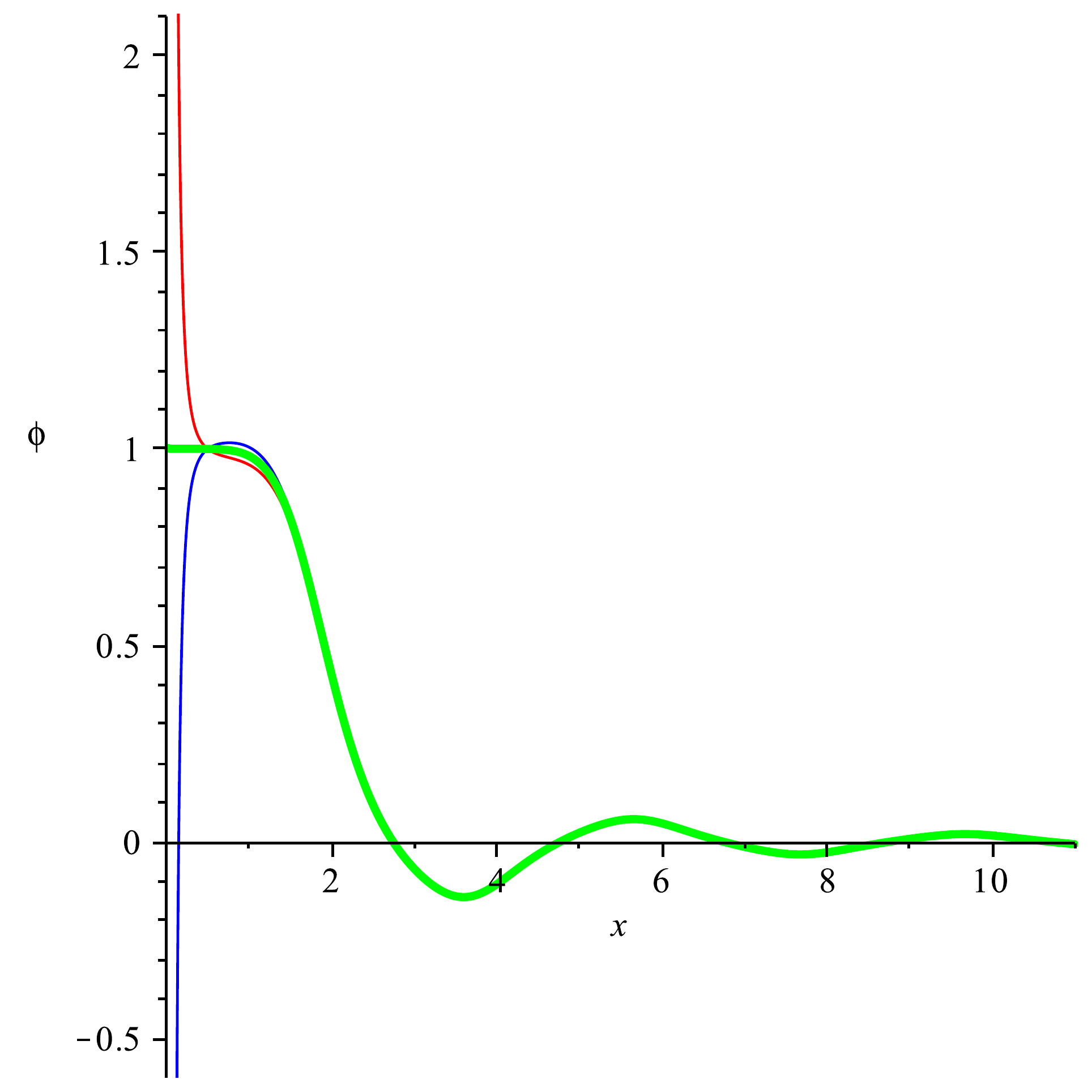, width=8cm}}
  \caption{Static spherically symmetric solutions $\phi(x)$ in non-linear CFT as a function of rescaled radius $x \equiv (3r)^{1/3}$.}
  \label{fig:static}
\end{figure}
\begin{figure}
  \centerline{\epsfig{file=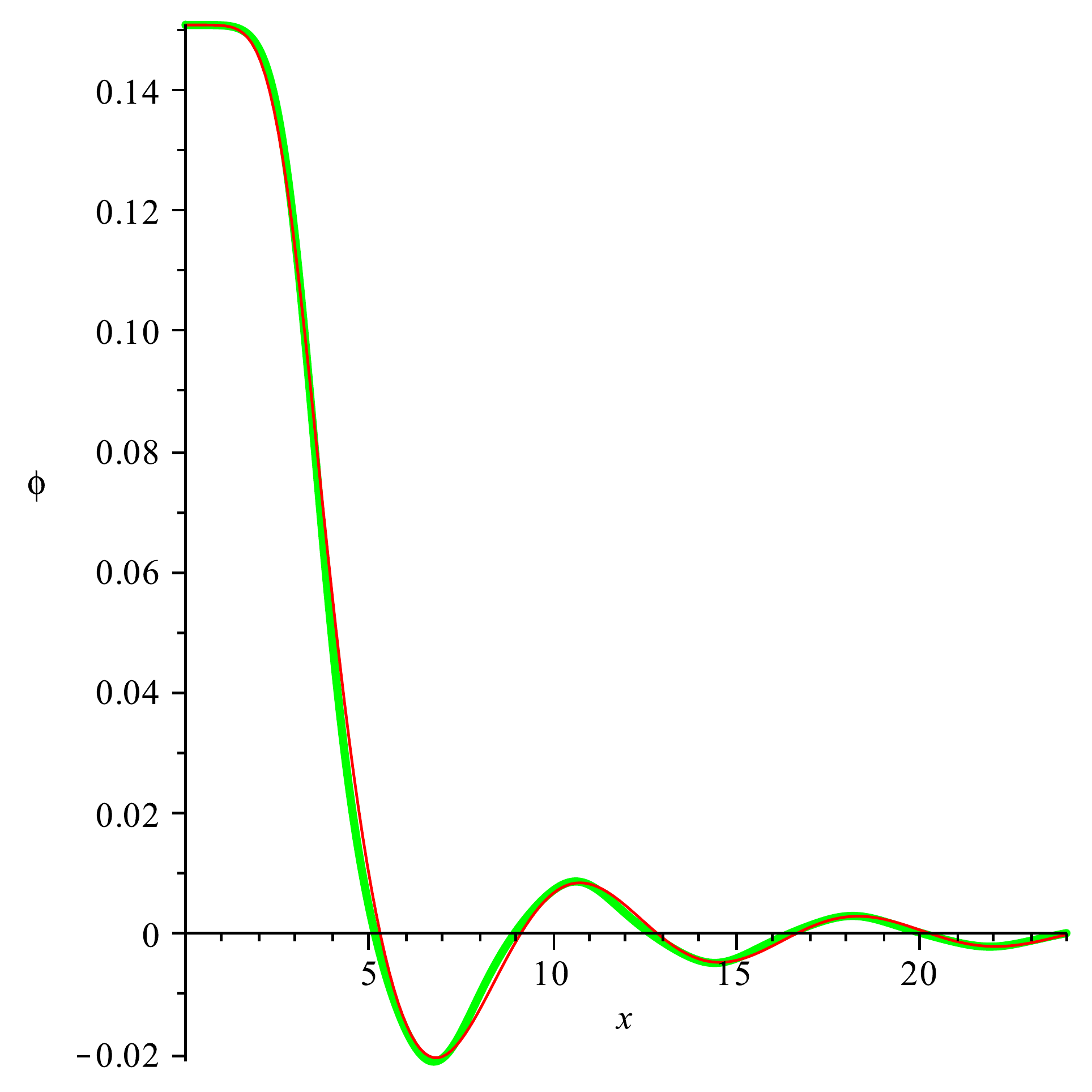, width=8cm}}
  \caption{Regular static spherically symmetric solution $\phi(x)$ in green versus global approximation (\ref{eq:static:approx}) in red.}
  \label{fig:approx}
\end{figure}

While Hamiltonian (\ref{eq:static:H}) is integrable, its integrals of motion are not easily obtainable in closed form \cite{Sarlet:1978, Struckmeier:2001}, so we find the solutions by numerical integration. Two initial conditions are required to specify the solution of a second order differential equation; one is the field value at origin $\phi_0$, the second is supplied by regularity condition (\ref{eq:static:regular}). Thus we have a one-parameter family of static everywhere regular solutions, which are localized around origin as shown in Fig.~\ref{fig:static}, and oscillate and decay away as $r^{-2/3}$ at spatial infinity.

A closer inspection of numerical results reveals that the regular solution transits from its behavior at origin to the limit cycle almost exactly like a cardinal function
\begin{equation}
  \psi_c(x) = x^{-1}\,\text{sn}(x,2^{-\frac{1}{2}}) - \text{cn}(x,2^{-\frac{1}{2}}).
\end{equation}
Swapping elliptical functions for trigonometric and introducing a few more coefficients to match power series at origin (\ref{eq:static:regular}), we obtain a global approximation of the regular solution in terms of elementary functions only
\begin{equation}\label{eq:static:approx}
  \psi(x) \simeq \left( \frac{\sin{kx}}{kx} - \cos{kx} \right) \sum\limits_{n=0}^4 a_n \frac{\sin(kx)^{2n}}{(kx)^{2n}},
\end{equation}
where coefficients $a_n$ are selected to keep only powers of $x^6$ in expansion around origin. As Fig.~\ref{fig:approx} shows, approximation (\ref{eq:static:approx}) reproduces the exact solution obtained by numerical integration remarkably well.

Solutions that blow up at origin require point-like external charges to source them, but existence of solitons implies that {\em the value of the source does not determine the solution uniquely}. To illustrate the point, Fig.~\ref{fig:static} also shows static solutions with positive and negative divergence at origin ``riding'' on top of the soliton and having virtually the same asymptotic away from the origin. Behavior of irregular solutions is substantially more complex then assumed in \cite{Dvali:2011uu}. As explained in Appendix~\ref{sec:running}, it involves {\em two distinct length scales}, not one.

\section{Dipole Bound States}
\label{sec:dipole}

\begin{figure}
  \epsfig{file=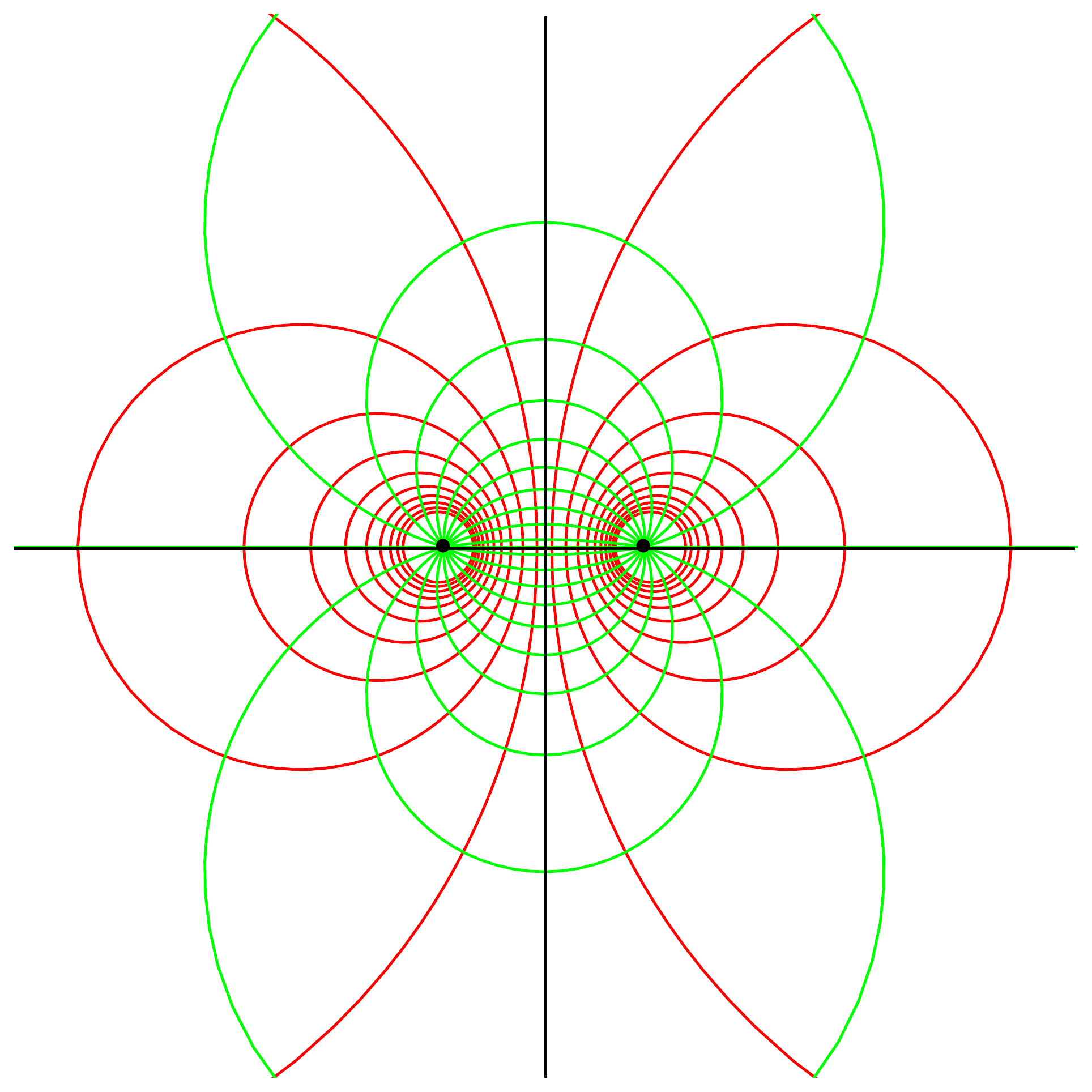, width=8cm}
  \caption{Bipolar coordinate system. Lines $\rho = \text{const}$ (shown in red) and $\theta = \text{const}$ (shown in green) are orthogonal.}
  \label{fig:bipolar}
\end{figure}

Although the solutions described above are everywhere regular and localized around the origin, the field tails decay too slowly ($\phi \sim r^{-2/3}$ at large $r$) for the total energy of the state to be finite. This is never a problem in practice as you always deal with finite truncations if you are trying to assemble the state dynamically, but is a reason we put a qualifier "soliton-like" instead of calling it a soliton. In this section, we show that in the soliton -- anti-soliton pair the slow-decaying tails interfere away, and the total energy is finite. Such a dipole pair forms an unstable bound state.

\begin{figure*}
  \begin{tabular}{cc}
    \epsfig{file=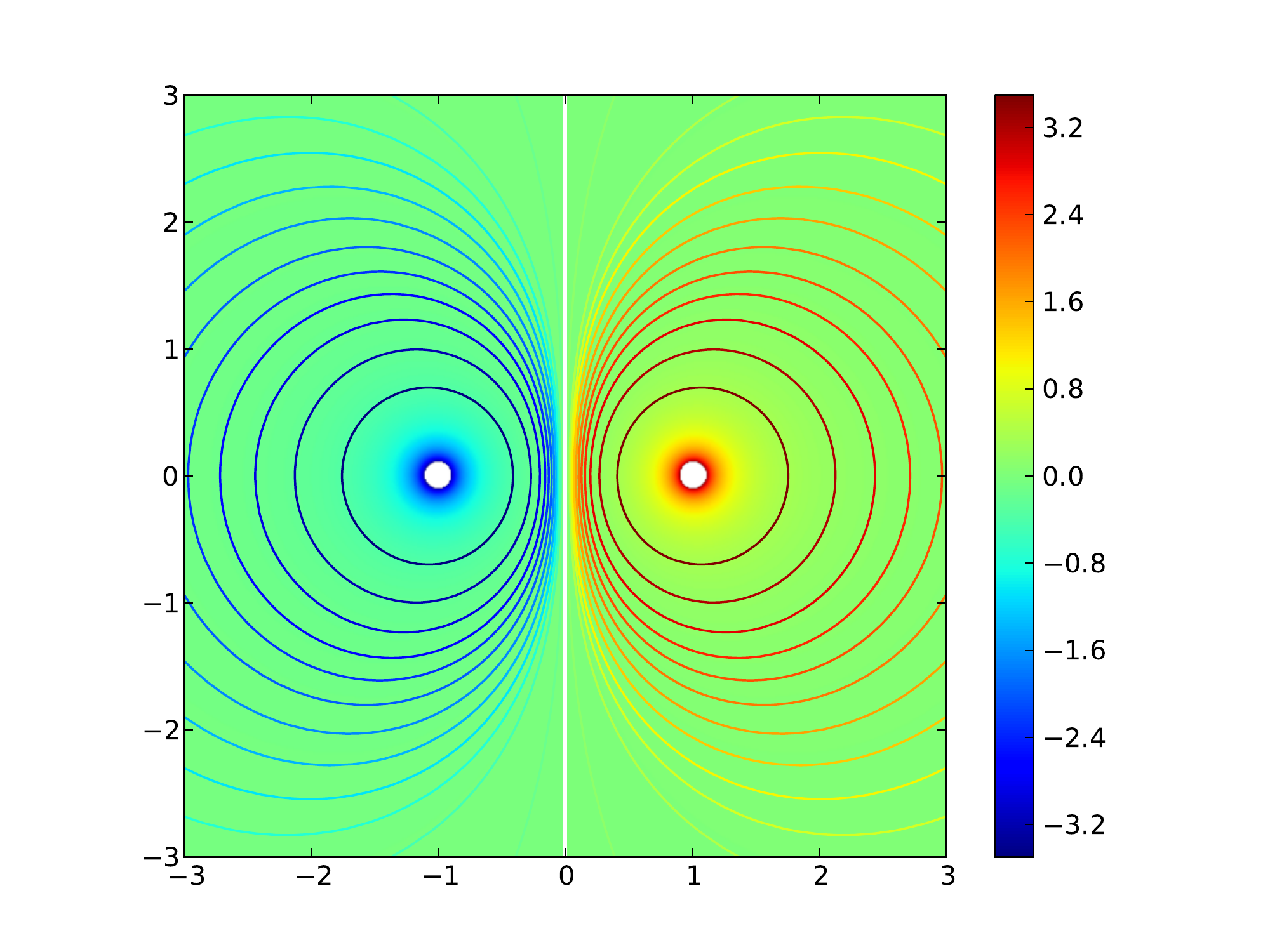, width=9cm} &
    \epsfig{file=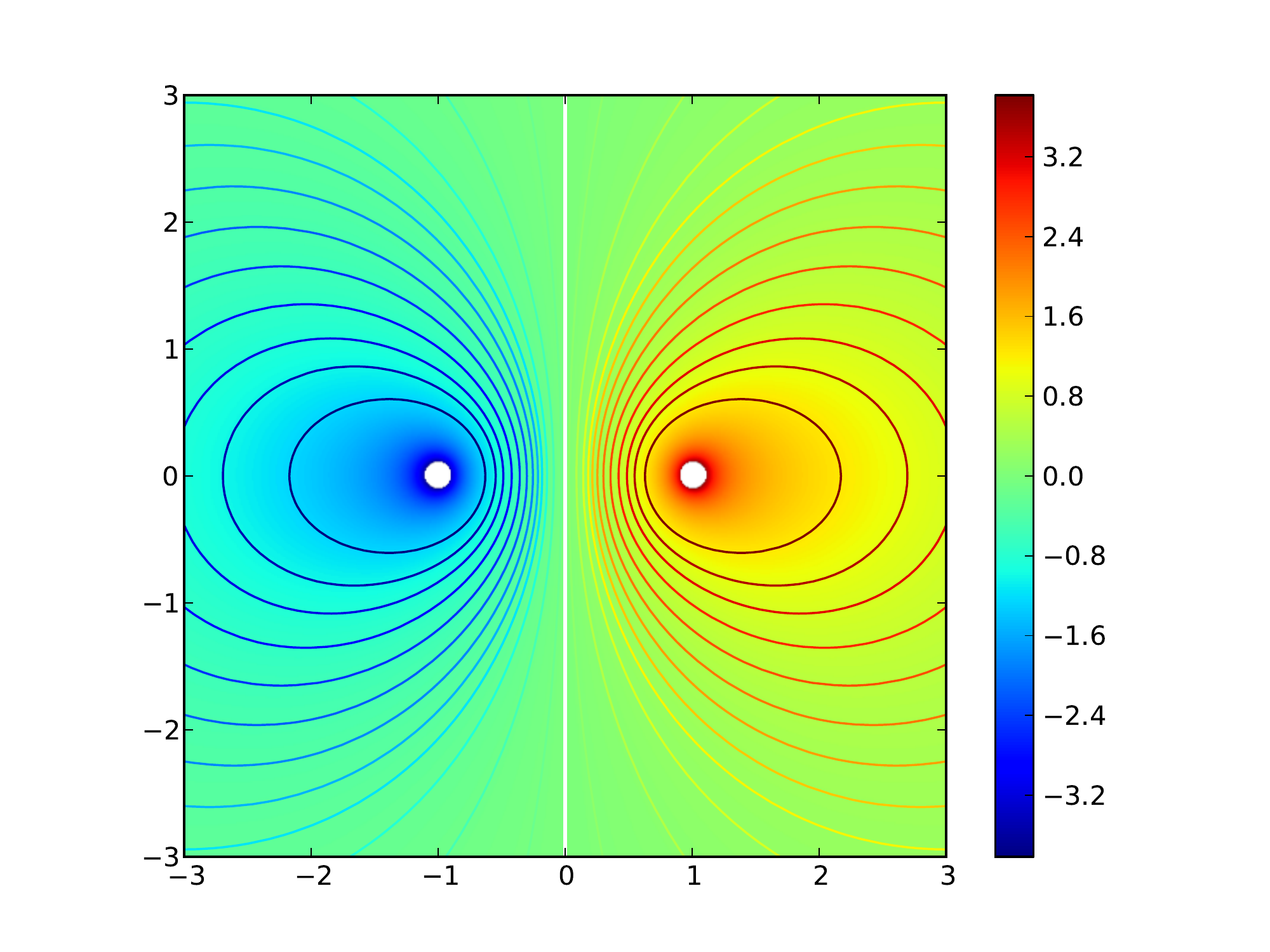, width=9cm} \\
    (a) $\lambda = 0$, $\phi_0 = 4$ &
    (b) $\lambda = 1$, $\phi_0 = 4$ \\
    \epsfig{file=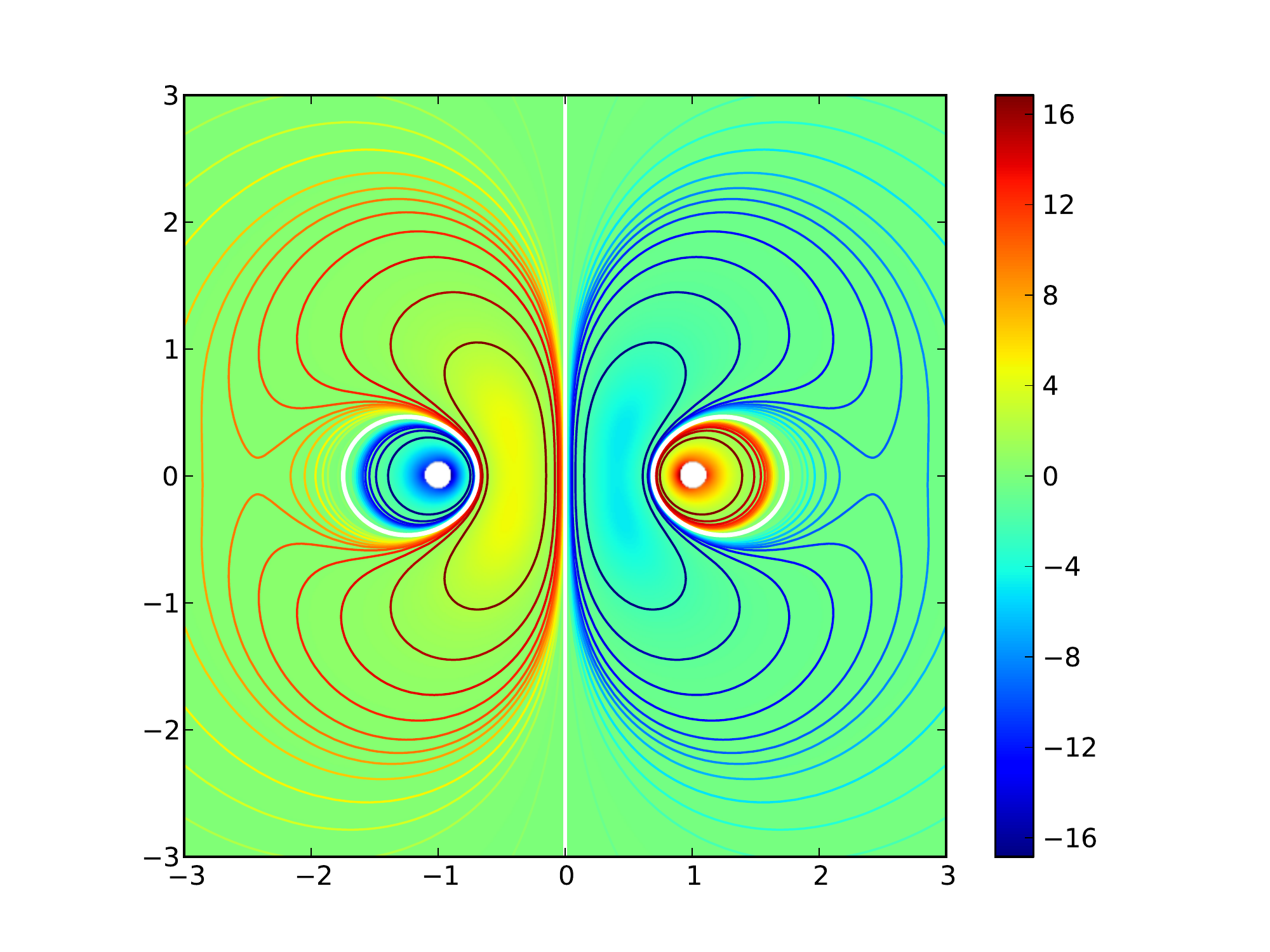, width=9cm} &
    \epsfig{file=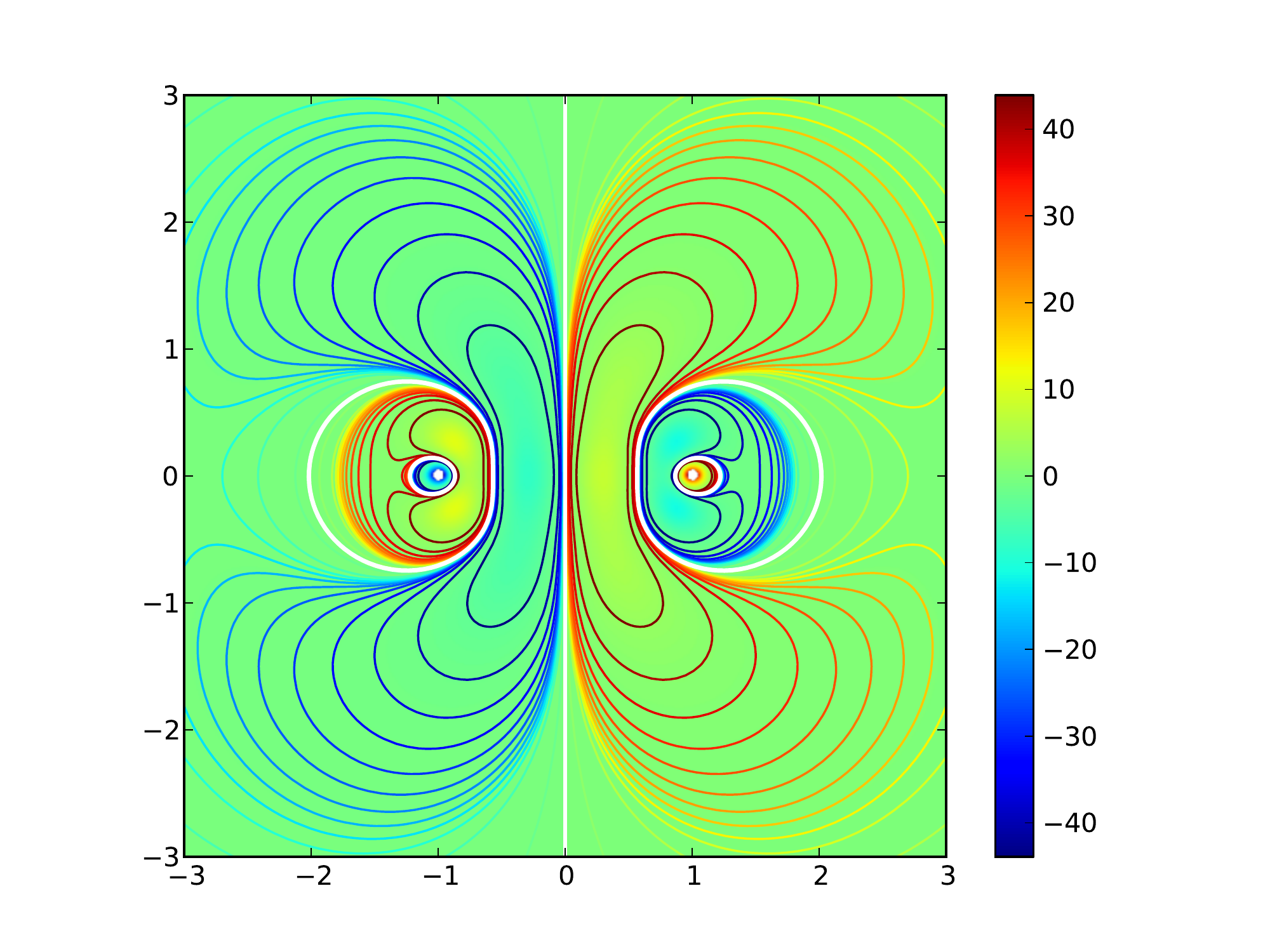, width=9cm} \\
    (c) $\lambda = 1$, $\phi_0 = 30$ &
    (d) $\lambda = 1$, $\phi_0 = 100$\\
  \end{tabular}
  \caption{Dipole solution in free theory (a), and a sequence of non-linear static solutions (b,c,d) as amplitude of the field around focal points is increased. Shaded contours show various isolines $\phi=\text{const}$, with thick white one highlighting $\phi=0$. Empty regions around focal points are excised from the solution domain, and might require charge distributions for regular completion.}
  \label{fig:dipole}
\end{figure*}

We will look for the static dipole solutions by spectral decomposition in bispherical coordinate system, which is adapted to the symmetries of the solution we seek \cite{Chaumet:1998}. Bispherical coordinates are obtained by axial rotation of a bipolar reference frame, which in turn is a conformal transformation of a $(\rho,\theta)$-plane by a factor
\begin{equation}
  \Omega = \cosh\rho - \cos\theta,
\end{equation}
which leaves coordinate grid orthogonal, but shrinks infinity $\rho=\pm\infty$ to two points -- the location of dipole foci -- as illustrated in Fig.~\ref{fig:bipolar}. Explicit transformation back to Cartesian coordinates reads
\begin{eqnarray}
  x &=& \Omega^{-1} \sinh\rho,\\
  y &=& \Omega^{-1} \sin\theta \cos\varphi,\nonumber\\
  z &=& \Omega^{-1} \sin\theta \sin\varphi,\nonumber
\end{eqnarray}
while the flat space metric in bispherical coordinates is
\begin{equation}
  ds^2 = \Omega^{-2} (d\rho^2 + d\theta^2 + \sin^2\theta\, d\varphi^2).
\end{equation}
Bispherical coordinate system allows separation of variables for Laplace equation, so we will decompose the solution we seek onto a countable set of basis functions
\begin{equation}\label{eq:decomp}
  \phi(\rho,\theta,\varphi) = \sum\limits_{k \ell m} u_{k \ell m} F_{k \ell m}(\rho,\theta,\varphi)
\end{equation}
formed by a tensor product of a radial basis $R_k(\rho)$ and the usual spherical harmonics $Y_{\ell m}(\theta,\varphi)$
\begin{equation}\label{eq:laplacian}
  F_{k \ell m}(\rho,\theta,\varphi) = (2\Omega)^{\frac{1}{2}} R_k(\rho)\, Y_{\ell m}(\theta,\varphi).
\end{equation}
With a little work, one can show that the variables in Laplace equation indeed separate, with
\begin{equation}
  \Delta F_{k \ell m} = 2^{\frac{1}{2}} \Omega^{\frac{5}{2}} \left[R_k'' - {\textstyle\frac{1}{4}} (2\ell+1)^2 R_k\right] Y_{\ell m}.
\end{equation}
The natural choice for radial basis in dipole symmetry is
\begin{equation}\label{eq:basis:Rk}
  R_k(\rho) = \sinh\left(k+{\textstyle\frac{1}{2}}\right)\rho,
\end{equation}
which makes the Laplace operator diagonal
\begin{equation}
  \Delta F_{k \ell m} = \Omega^2 \left[ \left(k+{\textstyle\frac{1}{2}}\right)^2 - \left(\ell+{\textstyle\frac{1}{2}}\right)^2 \right] F_{k \ell m},
\end{equation}
and the free field dipole solution formed by a pair of point charges $+q$ and $-q$ simply
\begin{equation}
  \phi_{\text{dipole}} = q F_{000}(\rho,\theta).
\end{equation}
However, as exponentials are rapidly growing functions as $\rho \rightarrow \pm\infty$, the convergence rate for coefficients of expansion in this basis is abysmal. Instead, we notice that basis functions (\ref{eq:basis:Rk}) are simply odd power polynomials in $\sinh\rho/2$, and recombine the basis set into Chebyshev polynomials of argument rescaled inside a bounded domain $|\rho| < \rho_{\text{max}}$
\begin{equation}
  \tilde{R}_k(\rho) = T_{2k+1}\left(\sigma\sinh\frac{\rho}{2}\right), \hspace{1em}
  \sigma^{-1} = \sinh\frac{\rho_{\text{max}}}{2}.
\end{equation}
The simplest way to evaluate the new radial basis is by using trigonometric representation
\begin{equation}
  \tilde{R}_k = \cos(n\varrho), \hspace{1em}
  \cos\varrho = \sigma\sinh\frac{\rho}{2}, \hspace{1em}
  n = 2k+1,
\end{equation}
while the second derivative of $\tilde{R}_k$ with respect to $\rho$ which enters the Laplace operator can be evaluated as
\begin{equation}
  \tilde{R}_k'' =
    \frac{1}{4} \frac{\sin n\varrho}{\sin^3 \varrho}\, n \left(1+\sigma^2\right) \cos\varrho\, -\,
    \frac{1}{4} \frac{\cos n\varrho}{\sin^2 \varrho}\, n^2 \left(\cos^2 \varrho +\sigma^2\right) .
\end{equation}
Evaluated at a set of points $\{\vec{x}_a\}$, Laplacian and field value operators (\ref{eq:decomp},\ref{eq:laplacian}) are just big matrices acting on a vector of spectral coefficients $\vec{u}$ drawn from a linear space spanned by values of composite index $k \ell m$ in (\ref{eq:decomp})
\begin{equation}
  (\Delta\phi)(\vec{x}) = \mathbb{D}\cdot \vec{u}, \hspace{1em}
  \phi(\vec{x}) = \mathbb{F} \cdot \vec{u}.
\end{equation}
Truncating the spectral expansion and selecting an appropriate collocation grid to evaluate the residuals, linear elliptical partial differential equation $\Delta\phi = f(x)$ turns into a linear algebra problem $\mathbb{D}\cdot \vec{u} = \vec{f}$ \cite{Boyd:2000}. An optimal grid is determined by the basis choice, which we take to be the Gauss-Lobatto grids for Chebyshev and Legendre polynomials in radial and angular directions, respectively
\begin{equation}
  \varrho_i = \frac{\pi}{2} \frac{N-i}{N-\frac{1}{2}}, \hspace{1em}
  P_{N-1}^2(\cos\theta_j) = 1.
\end{equation}
Direct inversion of the resulting linear algebra problem solves linear elliptic PDE to arbitrary precision \cite{Boyd:2000}. Alas, the same cannot be said even for semi-linear PDE.

It is well known in computational mathematics that elliptic semi-linear PDEs like $\Delta\phi+\phi^3=0$ admit infinitely many distinct solutions even in compact domains, even with trivial boundary conditions \cite{Chen:2004}. 
Without any additional constraints, positions and amplitude of the solitonic waves will be undetermined. To break this degeneracy, we impose explicit Dirichlet conditions on $\phi$ at arbitrary excision boundary $\rho=\rho_\text{max}$ close to focal points, and look for the solution outside. As all of the basis functions decay as $1/r^2$, no additional boundary conditions arise at spatial infinity ($\rho=0$, $\theta=0$). Then, starting with some initial guess, one tries to zero the residual
\begin{equation}
  b(x) \equiv \Delta\phi + \lambda\phi^3 - f(x) = 0
\end{equation}
by an iterative scheme. As one soon finds out, Newton-Kantarovich iteration obtained by linearizing the above equation
\begin{equation}
  \left(\mathbb{D} + 3\lambda\phi^2 \mathbb{F}\right) \cdot \delta \vec{u} = -\vec{b}
\end{equation}
fails to converge if non-linearity is sufficiently strong (and it is for oscillatory solutions we seek). So one is reduced to crawling down the direction of steepest descent in Levenberg-Marquardt fashion
\begin{equation}
  \left(\widetilde{\mathbb{L}^T\mathbb{L}}\right) \cdot \delta\vec{u} = -  \mathbb{L}^T \cdot \vec{B},
\end{equation}
and hoping for the best. Here tilde denotes the usual Levenberg-Marquardt regularization operator scaling the diagonal elements of a matrix by a factor of $1+\alpha$, and $\mathbb{L}$ and $\vec{B}$ are the preconditioned Newton iteration matrices
\begin{equation}
  \mathbb{L} = \Omega^{-\frac{3}{2}} \left(\mathbb{D} + 3\lambda\phi^2 \mathbb{F}\right), \hspace{1em}
  \vec{B} = \Omega^{-\frac{3}{2}}\,\vec{b}.
\end{equation}
Preconditioning by $\Omega^{-\frac{3}{2}}$ reflects the volume factors in the residual squared cost function
\begin{equation}
  \chi^2 = \int b^2(x)\, \sqrt{g\,}\,d^3x.
\end{equation}
Thus, after much effort and some applied magic having to do with initial guess, the solutions shown in Fig.~\ref{fig:dipole} were obtained. As degree of non-linearity is increased, the solution goes from a simple deformation (b) of a free dipole (a) through a sequence of sign reversals (c,d) which change the topology of $\phi=0$ surface. All of the dipole solutions have finite energy, as far away from the dipole field $\phi$ decays as $1/r^2$.

\section{Discussion}
\label{sec:discussion}

The most interesting result of this paper is that a version of critical collapse exists in a theory without gravity, and a simple one at that. The criticality is driven by an attractive  self-interaction of the field. The critical solutions are long-lived, and are likely the static soliton-like solutions we found. Inverse lifetime of a collapsing wavepacket scales linearly with amplitude for large amplitudes or homogeneous configurations, but switches over to a much shallower power law close to criticality. The value of critical exponent is determined by the spectrum of unstable perturbation of the critical solution \cite{Koike:1995jm}. This we leave for future study, but since approximate universality is observed in the critical collapse of arbitrary initial data, the spectrum should have a mass gap, or be discrete.

We further investigate static spherically symmetric solutions. The theory admits everywhere regular localized vacuum solutions -- solitons, but their energy is infinite due to slow decay of the oscillating tails at spatial infinity. Irregular solutions have an asymptotic at origin where effective coupling constant runs logarithmically, as the one in quantum chromodynamics. This, taken together with infinite energy of an isolated charge, lead to an interpretation of this theory as showing a classical analogue of confinement \cite{Dvali:2011uu}. Leaving aside the matter of solutions being unstable, we find the story is substantially more complicated --- in a general static solution two other intermediate asymptotic regimes interject between logarithmic running and the limit cycle contributing to energy divergence.

We also construct static non-linear solutions for soliton -- anti-soliton pair. Such a dipole has finite energy, and thus forms a bound state. As a final note, we must remark that the classical conformal field theory we consider here is deceptively simple, but actual non-linear solutions have very rich phenomenology. In many ways they are similar to solitonic solutions of Korteweg-de Vries equation and Petviashvili monopole vortices encountered in atmospheric science \cite{Tan:1997}. If one breaks conformal invariance by introducing mass terms, static solitons discussed here will probably turn into oscillating breather modes, not quite like in \cite{Segur:1987mg, Boyd:1990}, but rather related to oscillons discovered in \cite{Amin:2010jq, Amin:2010dc}.

\section*{Acknowledgments}\small
This work was supported by the Natural Sciences and Engineering Research Council of Canada under Discovery Grants program.

\appendix
\section{Classical Beta Function}
\label{sec:running}

\begin{figure}
  \epsfig{file=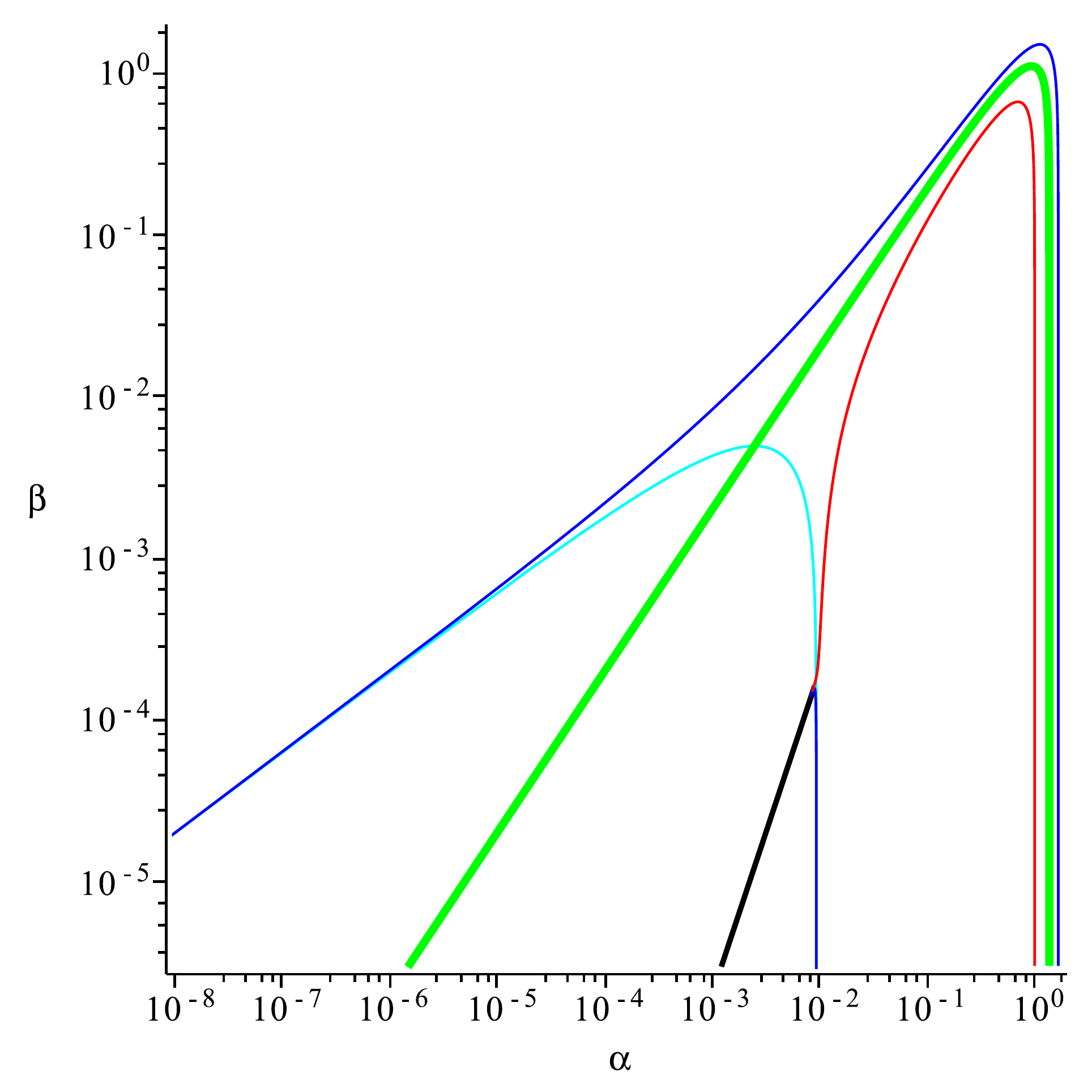, width=8cm}
  \caption{Classical renormalization group flow in terms of $\alpha \equiv \lambda q^2$ and $\beta \equiv d\alpha/d\ln r$ for static spherically symmetric solutions. Thick green line is a regular solution of Fig.~\ref{fig:static}, red and blue lines are irregular solutions with positive and negative divergence at origin, black line is asymptotic $\beta = 2\alpha^2$.}
  \label{fig:running}
\end{figure}

As the theory is scale-invariant, the equations of motion can be rewritten in an autonomous form. Introducing variable $q \equiv r\phi$ once again and switching over to logarithmic spatial scale $\mu\equiv\ln r$, equation (\ref{eq:static:phi}) becomes \cite{Dvali:2011uu}
\begin{equation}
  q'' - q' + \lambda q^3 = 0.
\end{equation}
Dynamical system can be analyzed in terms of classical renormalization group variables, effective coupling and its running
\begin{equation}
  \alpha(\mu) = \lambda q^2(\mu), \hspace{1em}
  \beta(\mu) = \alpha'(\mu).
\end{equation}
As with any autonomous dynamical system, the order can be reduced by considering momentum a function of generalized coordinate $\beta = \beta(\alpha)$, not the scale $\mu$
\begin{equation}\label{eq:rg}
  \beta' = \beta + \frac{\beta^2}{2\alpha}- 2 \alpha^2 = \beta\,\frac{d\beta}{d\alpha}.
\end{equation}
Behavior of solutions as $r \rightarrow 0$ corresponds to an UV fixed point at $\alpha=0$, but the equation (\ref{eq:rg}) admits {\em two} distinct flows to the fixed point, $\beta = 2\alpha$ and $\beta=2\alpha^2$ (and only two in the power law class). Substituting ansatz $\beta = c \alpha^\gamma$ in (\ref{eq:rg})
\begin{equation}
  \frac{c^2}{2} (1-2\gamma) \alpha^{2\gamma-1} + c \alpha^\gamma - 2\alpha^2 = 0,
\end{equation}
one can see that for leading terms to cancel as $\alpha \rightarrow 0$, the power exponent in the second term has to coincide with the first or the last one. First alternative yields $\beta = 2\alpha$, the second $\beta=2\alpha^2$. Integrating renormalization group equations give corresponding asymptotic solutions
\begin{eqnarray}
  \label{eq:rg:core}
  \alpha' = 2\alpha\phantom{^2} &\Longrightarrow&
  \alpha \equiv \lambda q^2 \propto r^2, ~
  \phi = \text{const}, \\
  \label{eq:rg:free}
  \alpha' = 2\alpha^2 &\Longrightarrow&
  \alpha^{-1} = 2\ln \frac{\ell_c}{r}, ~
  \phi \propto \frac{1}{r} \left(2 \ln \frac{\ell_c}{r}\right)^{-\frac{1}{2}}\!\!.
\end{eqnarray}
So, which one is chosen by the static solution? 

Turns out the general solution follows both asymptotics in turn, abruptly switching over at a certain scale $\ell_*$, as illustrated in Fig.~\ref{fig:running}. This can happen because, while the theory is scale-invariant, {\em the solutions are not}. Regular soliton solutions are characterized by a single length scale $\ell_0 = \lambda^{-1/2}\phi_0^{-1}$. Irregular solutions introduce a second length scale $\ell_q = q\phi_0^{-1}$ related to the value of charge sourcing them. Starting from $r=\infty$ to $r=0$, the general solution then goes through four distinct asymptotic regions: the limit cycle (\ref{eq:cn}), the soliton core (\ref{eq:rg:core}), the transition over $q\simeq\text{const}$ branch, and finally the asymptotically free regime with logarithmic running (\ref{eq:rg:free}).

\begin{thebibliography}{99}

\bibitem{Aharony:1999ti}
  O.~Aharony, S.~S.~Gubser, J.~M.~Maldacena, H.~Ooguri, Y.~Oz,
  {\it ``Large $N$ field theories, string theory and gravity,''}
  Phys.\ Rept.\  {\bf 323}, 183-386 (2000)
  [\arXiv{hep-th/9905111}].

\bibitem{Symanzik:1973hx}
  K.~Symanzik,
  {\it ``A field theory with computable large-momenta behavior,''}
  Lett.\ Nuovo Cim.\  {\bf 6S2}, 77-80 (1973).

\bibitem{Parisi:1973ma}
  G.~Parisi,
  {\it ``Deep inelastic scattering in a field theory with computable large-momenta behaviour,''}
  Lett.\ Nuovo Cim.\  {\bf 7S2}, 84-88 (1973).

\bibitem{Kleinert:1991rg}
  H.~Kleinert, J.~Neu, V.~Schulte-Frohlinde, K.~G.~Chetyrkin, S.~A.~Larin,
  {\it ``Five loop renormalization group functions of $O(n)$ symmetric $\phi^4$ theory and $\epsilon$-expansions of critical exponents up to $\epsilon^5$,''}
  Phys.\ Lett.\  {\bf B272}, 39-44 (1991)
  [\arXiv{hep-th/9503230}].

\bibitem{Dvali:2011uu}
  G.~Dvali, C.~Gomez, S.~Mukhanov,
  {\it ``Classical dimensional transmutation and confinement,''}
  \arXiv[hep-th]{1107.0870}.

\bibitem{Rubakov:2009np}
  V.~A.~Rubakov,
  {\it ``Harrison-Zeldovich spectrum from conformal invariance,''}
  JCAP {\bf 0909}, 030 (2009)
  [\arXiv[hep-th]{0906.3693}].

\bibitem{Libanov:2010nk}
  M.~Libanov, V.~Rubakov,
  {\it ``Cosmological density perturbations from conformal scalar field: infrared properties and statistical anisotropy,''}
  JCAP {\bf 1011}, 045 (2010)
  [\arXiv[hep-th]{1007.4949}].

\bibitem{Choptuik:1992jv}
  M.~W.~Choptuik,
  {\it ``Universality and scaling in gravitational collapse of a massless scalar field,''}
  Phys.\ Rev.\ Lett.\  {\bf 70}, 9-12 (1993).

\bibitem{Koike:1995jm}
  T.~Koike, T.~Hara and S.~Adachi,
  {\it ``Critical behavior in gravitational collapse of radiation fluid,''}
  Phys.\ Rev.\ Lett.\  {\bf 74}, 5170 (1995)
  [\arXiv{gr-qc/9503007}].

\bibitem{Gundlach:2007gc}
  C.~Gundlach, J.~M.~Martin-Garcia,
  {\it ``Critical phenomena in gravitational collapse,''}
  Living Rev.\ Rel.\  {\bf 10}, 5 (2007)
  [\arXiv[gr-qc]{0711.4620}].

\bibitem{Kiper:1984}
  A.~Kiper,
  {\it ``Fourier series coefficients for powers of the Jacobian elliptic functions,''}
  Mathematics of Computation {\bf 43}, 247-259 (1984).

\bibitem{Sarlet:1978}
  W.~Sarlet ,
  {\it ``Exact invariants for time-dependent Hamiltonian systems with one degree-of-freedom,''}
  J.\ Phys.\ A: Math.\ Gen.\  {bf 11}, 843 (1978).

\bibitem{Struckmeier:2001}
  J.~Struckmeier and C.~Riedel, 
  {\it ``Invariants for time-dependent Hamiltonian systems,''}
  Phys.\ Rev.\ E  {\bf 64}, 026503 (2001).

\bibitem{Chaumet:1998}
  P.~Chaumet,
  {\it ``Electric potential and field between two different spheres,''}
  Journal of Electrostatics {\bf 43}, 145-159 (1998).

\bibitem{Boyd:2000}
  J.~P.~Boyd
  {\it ``Chebyshev and Fourier spectral methods,''}
  Dover Publications; 2nd revised edition (2000).

\bibitem{Chen:2004}
  C.~Chen, Z.~Xie,
  {\it ``Search extension method for multiple solutions of a nonlinear problem,''}
  Computers \& Mathematics with Applications, {\bf 47} 327-343 (2004).

\bibitem{Tan:1997}
  B.~Tan and J.~P.~Boyd,
  {\it ``Dynamics of the Flierl-Petviashvili monopoles in a barotropic model with topographic forcing,''}
  Wave Motion {\bf 26}, 239-251 (1997).

\bibitem{Segur:1987mg}
  H.~Segur, M.~D.~Kruskal,
  {\it ``Nonexistence of small amplitude breather solutions in $\phi^4$ theory,''}
  Phys.\ Rev.\ Lett.\  {\bf 58}, 747-750 (1987).

\bibitem{Boyd:1990}
  J.~P.~Boyd,
  {\it ``A numerical calculation of a weakly non-local solitary wave: the $\phi^4$ breather,''}
  Nonlinearity {\bf 3}, 177 (1990)

\bibitem{Amin:2010jq}
  M.~A.~Amin, D.~Shirokoff,
  {\it ``Flat-top oscillons in an expanding universe,''}
  Phys.\ Rev.\  {\bf D81}, 085045 (2010)
  [\arXiv[astro-ph.CO]{1002.3380}].

\bibitem{Amin:2010dc}
  M.~A.~Amin, R.~Easther, H.~Finkel,
  {\it ``Inflaton fragmentation and oscillon formation in three dimensions,''}
  JCAP {\bf 1012}, 001 (2010)
  [\arXiv[astro-ph.CO]{1009.2505}]

\end{thebibliography}

\end{document}